\documentclass[smallextended]{svjour3}  
\usepackage{wrapfig}
\usepackage{times}
\usepackage{graphicx}
\usepackage{amssymb,amsmath}
\journalname{Empirical Software Engineering}
\begin{document}
\title{Inferencing into the void: problems with implicit populations}
\subtitle{Comments on `Empirical software engineering experts on the use of students and professionals in experiments'}
\author{Martin Shepperd}

\institute{M. Shepperd \at
              Brunel University London, UK \\
              \email{martin.shepperd@brunel.ac.uk} 
}

\date{Received: 15th January 2018 / Revised 25th April 2018 / Accepted: 15th October 2018}

\maketitle

\section{Introduction} \label{Sec:Intro}
I welcome the contribution from Falessi et al.~\cite{Fale18} hereafter referred to as F++ , and the ensuing debate.  Experimentation is an important tool within empirical software engineering, so how we select participants\footnote{Note that the focus is on human-centric experiments so computational experiments (e.g., comparing different machine learners to predict software defects) are excluded.} is clearly a relevant question.  Moreover as F++ point out, the question is considerably more nuanced than the simple dichotomy it might appear to be at first sight.

The majority of software engineering experiments to date use students as participants.  The 2005 systematic review of Sj{\o}berg et al.~\cite{Sjob05} found that less than 10\% of experiments used professionals.  More recently daSilva et al.~\cite{daSi12} found in a mapping study of replication studies that 59\% used students / researchers, 15\% used a mix and 12\% used solely professionals as participants.  Ko et al.~\cite{Ko15} report a decline from a peak of 44\% of tool evaluations using at least one professional participant in  2002 to 26\% in 2011.  They also reported that an astonishing 23\% of studies failed to report any information concerning the nature of their participants.  So it seems clear that researchers still predominantly use student participants.

Our discipline is entitled software \emph{engineering}.  Many of our concerns and challenges relate to scale.  Additionally, engineers must work in complex, dynamic and imperfectly understood environments.  How representative are students (i.e., how strong is the external validity) and does this matter?  Is their use a valid design decision, a pragmatic compromise, or positively misleading?  

The remainder of this commentary is structured as follows.  In Section \ref{Sec:F++Summary} I briefly summarise the arguments of F++ and comment on their approach.  Next, in Section \ref{Sec:Principles}, I take a step back to consider the nature of representativeness in inferential arguments and the need for careful definition.  Then I give three examples of using different types of participant to consider impact.  I conclude by arguing --- largely in agreement with F++ --- that the question of whether student participants are representative or not depends on the target population. However, we need to give careful consideration to defining that population and, in particular, not to overlook the representativeness of \emph{tasks} and \emph{environment}.  This is facilitated by explicit description of the target populations.

\section{Summary of F++} \label{Sec:F++Summary}

The objective of F++ is principally to a obtain deeper understanding of the representativeness of software engineering experiments conducted using student participants.  This is key to consideration of generalisabiity.  In addressing this challenge they make two observations. First, that there may be circumstances when students behave similarly to professionals. Second, students and professionals are not two dichotomous classes, since a student may have prior professional experience, might be working part-time, might be woking as an intern, or his or her educational experience might be highly relevant or realistic to phenomena under investigation.  Conversely a professional might have limited experience, indeed could theoretically be in transition (i.e., yesterday a student, today an employee).  It is difficult to disagree.

They then obtained further evidence via a large focus group of 65 `experts' and followed it up with a survey.  However, one cannot help being concerned that not all conclusions strictly follow.  For example, the fact that one author has had poor experiences using professionals in experiments with low numbers and high drop out rates and difficulties getting them to adhere to certain techniques does not mean this must be so.  By contrast, I, Carolyn Mair and Magne J{\o}rgensen have recently completed a series of experiments with more than 400 professional participants \cite{Shep18}.  Clearly the challenges vary, but equally clearly it's not impossible to conduct experiments with large sample sizes.  Additionally, if unwillingness to adhere to a particular technique or intervention differs between students and professionals, this would seem to weaken the value of students as proxies.

The proposal to improve the description of subjects beyond merely whether they are professionals or not is useful.  However,  limiting it to a 3-class system of real, relevant and recent seems generally overly restrictive.  Again context is all, but minimally one might try to use experience as a continuous variable along with other factors such as role, seniority,  and so forth.

A parenthetic concern arises from the stated motivation that ``we have observed too many times that our papers were rejected because we used students as subjects É" \cite{Fale18}.  I fear this betrays a driving force behind some aspects of F++'s paper, despite the difficulty that it's unknowable whether their (or anybody else's) papers have been incorrectly rejected.

Moving on, as F++ remark, ``every subject sample is representative of a certain population".  For meaningful discussion, however, it's important to define the population.  Without this, discussion of sampling and representativeness is indeterminate.  Consequently, the remainder of this commentary reflects on the ideas of sampling, representativeness and the population into which the researchers are inferring.

\section{First principles} \label{Sec:Principles}

So the question is what do we mean by representativeness and how might this concept be operationalised?  Wieringa and Daneva \cite{Wier15} refer to software engineering experimentation as a sample-based lab-to-field strategy of generalisation where research should be aimed at improving the accuracy of claims regarding scope (i.e., the set of phenomena to which the claims apply given the presently-available arguments and evidence).  

The reasoning process is one of inference since we wish the scope of ideas or theories to apply more widely than just those that have been experimentally observed.  That is we use inference to generalise.  In the experimental paradigm we are most familiar with statistical inference.  The sample data are used to generalise statistically to a \emph{well-defined} population, called the study population where the sample is a subset of the study population.  The sample should be probabilistic, i.e., we know the likelihood of any member being selected.  Often we might wish to extend the generalisation beyond the study population to a ``theoretical population", of which the study population is a subset \cite{Wier15}. The theoretical population may be less well-defined than the study population, in terms of a sampling frame so it might be possible to have a list of all software engineers at Company X, but unlikely to have one for all software engineers globally.  

Wilkinson and the Task Force on Statistical Inference \cite{Wilk99} point out that sometimes ``the case for the representativeness of a convenience sample can be strengthened by explicit comparison of sample characteristics with those of a defined population across a wide range of variables".  All this requires not only a thorough description of contextual factors but also explicit consideration of the population, or in Wieringa and Daneva's terminology \cite{Wier15} the scope.  If we don't know into what population we're inferencing then it's hard to see how the process can be open to meaningful scrutiny let alone be considered rigorous.  Unfortunately, it does not seem to be common practice in empirical software engineering to formally describe the population even if the inferential statistics are described in considerable detail. 

This informality also leads to an under-appreciation of the fact that in software engineering the population is not restricted to the human participants (be they professional or otherwise) but also tasks in particular settings.  In other words, most experiments are not only concerned with applying different treatments to different participants, e.g., some might use test-driven development and others traditional development, but this also needs to be applied to particular artefacts in a particular environment.  When we infer from the experimental results, we want to generalise to more than just one software artefact and in more than one setting.

This implicit view of a population leads to two further difficulties.  First, it is hard to judge the quality of the sample.  And although most software engineering samples are are far from probabilistic --- be they students or professionals --- which considerably undermines any statistical generalisation, some form of analogical inferencing remains possible \cite{Rict13,Wier15} which could augment the generalisation arguments.

\section{Examples}


An interesting example, drawn from experimental economics \cite{Levi10}, used real-world experts in a laboratory setting where they were asked to engage in 2-player games to investigate the ubiquity of minimax strategies.  However, the authors found different behaviour in the lab from the participants' professional settings (e.g. as professional poker players) and therefore concluded that it's not just the participant but the context and task as well.  It is easy to envisage similar situations in software engineering.

A second example, is based on group estimation and Boehm's delphi estimation process. In this experiment \cite{Pass03}, we used Master's students as proxies for professionals.  The estimation task we gave them was relatively trivial and the context was a laboratory rather than the complexities and uncertainties deriving from a large organisational setting.  In terms of representativeness or external validity there were considerable weaknesses and this substantially undermined our ability to generalise.  The experimental design decisions were solely for practical reasons.  As a pilot study the work may have had some value.  As means of saying much about interesting software engineering settings it was decidedly limited.

A third example, was an experiment by Salman et al.~\cite{Salm15} [and cited by F++] to explicitly compare behaviours of professionals and students when using test-driven development methods.  Interestingly, they conducted the experiment with students in an academic setting and with professionals in a commercial setting.  They found both groups performed relatively similarly when in an initial learning situation, but there were other differences such as professionals producing larger but less complex solutions.  This suggests that professionals and students need not always differ, nor need they always behave in similar ways.  And we won't know unless at least some experiments or studies use professionals.  Furthermore, it is unclear to what extent the differences are due to participants, tasks or the setting, or an interaction between all three.

The point from all three examples is that we should consider the representativeness of participants, tasks \emph{and} setting.  Equally important we need to be aware of how they \emph{interact}.

\section{Summary} \label{Sec:Summary}

Our concern has been experiments, but we really need a broader set of research methods, e.g., simulation, case studies and action research in order to address realism and the practicality that an engineering discipline demands.  Also there are alternatives to statistical inference and in particular generalisation by analogy \cite{Wier15}.  These should be further explored given our almost invariable use of non-statistical sampling.

The discussion thus far has focused on representativeness, but there is also the point that if we want our research to influence practice it is likely to be more influential when we do not use proxies, however representative.  In other words, there is also a `marketing' or communication to practitioners angle.  Use of at least some professional participants undertaking realistic tasks in realistic settings is therefore important if we want our research to have impact.

So to conclude, demanding perfection in the design and execution of our experiments is a counsel of despair; there is clear value in using students especially for preliminary experiments.  But if the target population is some class of professional, it is hard to see why, if practically possible, it would be disadvantageous to use professional participants. The likelihood that this might entail more effort should not be a reason for not undertaking such empirical work.  But underpinning all other issues, more consideration needs to be given to how we sample tasks and environments as well, and how these aspects of our experiments reflect professional `reality'.  These must be addressed explicitly.  Otherwise we are in danger of inferencing into an intellectual void.

\bibliography{EMSE_commentary}
\bibliographystyle{spmpsci}  

\begin{acknowledgements}
I would like to thank the editors of the Empirical Software Engineering journal, Robert Feldt and Tom Zimmerman for the opportunity to publicly contribute to the debate on the choice of participants in experimentation.  I'm also grateful to Steve Counsell and Nour Ali for their careful reading and helpful suggestions on an earlier version of this commentary.
\end{acknowledgements}

\end{document}